\documentclass[10pt,twoside]{article}
\usepackage{graphicx}
\usepackage{amsmath,amssymb}
\usepackage[mathscr]{eucal}
\def\hang{\hangindent\parindent}
\def\textindent#1{\indent\llap{#1\enspace}\ignorespaces}

\begin{document}

\centerline{\Large\bf Synchronization of Limit Sets
 }
 \vspace{0.3cm}

\vskip 3pt \centerline{\bf C.P. Li$^{\,1}$ and W.H.
Deng$^{\,2,1}$}

\vskip 3pt

\centerline{\footnotesize\it $^1$Department of Mathematics,
Shanghai University, Shanghai 200444,
 China}

\centerline{\footnotesize\it $^2$School of Mathematics and
Statistics, Lanzhou University, Lanzhou 730000, China}

\begin{center}
\begin{minipage}{12cm}{\small\quad In this Letter, we derive a sufficient
condition of synchronizing limit sets (attractors and repellers)
by using the linear feedback control technique proposed here.
There examples are included. The numerical simulations and
computer graphics show that our method work well. }
\end{minipage}
\end{center}

Historically, the study of synchronization phenomena of dynamical
systems has been an active topic in physics.
 In the 17th century, Huggens found two synchronization
 clocks, other early discovered examples, such as, wobbly bridges,
 the oscillating uniformly Josephson junctions, the synchronized
 lightning fireflies, synchronization of adjacent organ pipes, emerging
 coherence in chemical oscillators, etc [1].

 In recent decades, the research on synchronization moved to
 chaotic systems. As we know, an essential characteristic of chaotic
 system is that its evolution sensitively depends on initial conditions,
 intuitionally, this intrinsically defy synchronization. But in 1990 Pecora and
 Carroll [2] pointed out that when we coupled two identical chaotic systems,
 the synchronization between them is possible. In the sequel, lots of methods
 and techniques of synchronizing two chaotic systems were proposed
 and studied. Amongst these methods, one important method is the linear feedback method due
 to the fact that the drive and response systems become weakly coupled in the
 process of synchronization and this can easily be implemented in circuits. As far
 as we know, there have been no theoretical results available for some interesting
 chaotic systems to ensure that these systems can be synchronized by using the usual
 linear feedback technique, for example, R\"{o}ssler system. Although
 it just has one nonlinear term, the R\"{o}ssler system doesn't have symmetry property
 (Lorenz system has two nonlinear terms, but it has symmetry
 properties), so it is usually difficult to construct a Lyapunov function
 for proving the global asymptotical stability of the error system.
 In this Letter, we take the R\"{o}ssler system as an example. The associate
 synchronization is easily realized by utilizing our method.

 On the other hand, when the conditional Lyapunov exponents of two coupled
 systems are all negative, then it is usually thought that these two systems
 can be synchronized [3]. However, it has been recently reported that the
 negativity of conditional Lyapunov exponents is neither a
 sufficient condition nor a necessary condition for chaos
 synchronization because of some unstable invariant sets in the
 stable synchronization manifold, see Refs. [4] and [5], and the
 references cited therein. This motivates us to find a sufficient
 even universal condition of synchronization for more chaotic
 systems. In this Letter, we propose a sufficient condition of
 synchronization of limit sets. It is known that present studies of synchronization
 are for chaotic attractors (i.e., stable limit sets). Unlike current
 research, in our article, the limit sets are not necessary to be {\bf stable limit
 sets} (attractors), they even are {\bf unstable limit sets} (repellers), for example,
 unstable limit cycle. As far as we know, this is the first time
 to consider synchronization of unstable limit sets.

Consider the following system
$$
\frac{dX}{dt}=f(X), \eqno(1)
$$
where  $X=(x_1,\,x_2,\,\cdots,\,x_n)^T \in \mathbb{R}^n$, $f:\,
\mathbb{R}^n \rightarrow \mathbb{R}^n$ is differentiable.

We build the drive and response systems as follows, respectively,
$$
\frac{dX}{dt}=f(x_1,\,\cdots,\,x_k,\,x_{k+1},\,\cdots,\,x_n),
\eqno(2)
$$
and
$$
\frac{dY}{dt}=f(x_1,\,\cdots,\,x_k,\,y_{k+1},\,\cdots,\,y_n)+u(Y-X),
\eqno(3)
$$
in which  $Y=(y_1,\,y_2,\,\cdots,\,y_n)^T,\,
\tilde{Y}=(x_1,\,\cdots,\,x_k,\,y_{k+1},\,\cdots,\,y_n)^T\in
\mathbb{R}^n$, $ 0 \le k<n$, $u$ is the control parameter. Here
$k=0$ corresponds to one way coupled synchronization approach
[6,7].

Letting $E=Y-X$, $\tilde{E}=\tilde{Y}-X$, and subtracting (2) from
(3) yield [8,9]
$$
\begin{array}{lll}
\displaystyle\frac{{\rm d}\,E}{{\rm d}\,t}&=& f(\tilde{Y})-f(X)+uE \\
                        &=& \int_0^1
\displaystyle  Df(X+s\tilde{E})\tilde{E}\,{\rm d}\,s+uE
\end{array}
$$
Now, we choose a  Lyapunov function as $V=\displaystyle
\|E\|^2/2$, then,
$$
\begin{array}{lll}
\displaystyle\frac{{\rm d}\,V}{{\rm d}\,t}=<\frac{{\rm d}E}{{\rm
d}t},E>& = & <\int_0^1
\displaystyle Df(X+s\tilde{E})\tilde{E}{\rm d}\,s, E > +u\|E\|^2 \\
& \le & \| Df \| \cdot \| \tilde{E} \| \cdot \| E \|+u \|E\|^2 \\
& \le & (\| Df \|+u) \|E\|^2,
\end{array}
$$
where $ \| Df \|$ denotes $\sup\limits_{X \in \mathbb{R}^n} \|
Df(X) \| $, vector norm is 2-norm, matrix norm is the spectral
norm.

So, if $\| Df(X) \|$ is bounded by a constant $M$, i.e., $\| Df
\|<M$, then we can choose $u<-M$ such that synchronization between
systems (2) and (3) can be reached. In more details, if system (2)
has a (stable or unstable) limit $\Omega$, an long as we choose $
u<-M=-\sup\limits_{X \in \Omega} \| Df(X) \| $, then this limit
set can be synchronized. Till now, almost all publications are for
synchronization of stable limit sets, for example, chaos
synchronization. But synchronization of unstable limit sets has
not been studied yet. This paper is the first one to consider such
a topic. On the other hand, for any continuously differentiable
system, any (stable and unstable) limit sets of this system can be
synchronized with the aid of a simple linear feedback controller
derived here. In this sense, our derived synchronization method is
universal.

It is evident that the condition of synchronization in this
article is sufficient but not necessary. Generally speaking, to
estimate $\sup\limits_{X \in \Omega} \| Df(X) \|$ is not easy due
to two facts: 1) the bound of the limit set $\Omega$ is often
difficult to estimate, 2) the corresponding eigenvalues of
$(Df(X))^TDf(X)$ are difficult to determine. So we can find a
suitable $u$ by numerical exploration such that synchronization
can be realized. This numerical exploration is described as
follows, $1^0)$ choose a suitably large $-u$ such that
synchronization can be realized, $2^0)$ check $u/2$ whether or not
synchronization can be realized, if does, repeat $2^0)$;
otherwise, choose $3u/4$ and repeat $2^0)$. This process can not
be ended until the control parameter $\mid u\mid$ is suitable
small. Generally speaking, choosing a small $\mid u\mid$ is
economical.

We firstly consider synchronization of a unstable limit set.

It is known that Lorenz system [10]
$$
\displaystyle\frac{dx}{dt}=\sigma(y-x),~~\frac{dy}{dt}=rx-y-xz,~~~~\frac{dz}{dt}=xy-bz,\eqno(4)
$$
has a unstable limit cycle $\mathcal{C}$ (a repeller) if
$\sigma=10$, $b=8/3$, $r=24.5$. See Fig. 1. The drive-response
system is constructed as follows, respectively,
$$
\displaystyle\frac{dx_m}{dt}=\sigma(y_m-x_m),~~\frac{dy_m}{dt}=rx_m-y_m-x_mz_m,~~\frac{dz_m}{dt}=x_my_m-b_mz_m,
$$
and
$$
\displaystyle\frac{dx_s}{dt}=\sigma(y_s-x_s)+u(x_s-x_m),~~\frac{dy_s}{dt}=rx_s-y_s-x_sz_s+u(y_s-y_m),
~~\frac{dz_s}{dt}=x_sy_s-b_sz_s+u(x_s-z_m).
$$
The simulations are displayed in Fig. 1, where $u=-6$ is chosen by
numerical exploration.

For a (chaotic) system, the drive signal can not be randomly
chosen, otherwise, the expected synchronization can not be
implemented. For example, in Lorenz system, if we define variable
$z$ as the drive signal, the rest as response signals,
synchronization between the drive and response systems can not be
realized by using the usual (Pecora-Carroll) method, for details,
see p. 7 of Ref. [6]. Here we can realize synchronization by
adding a simple controller. The drive-response system is
constructed below.
$$
\displaystyle\frac{dx_m}{dt}=\sigma(y_m-x_m),~~\frac{dy_m}{dt}=rx_m-y_m-x_mz_m,~~\frac{dz_m}{dt}=x_my_m-b_mz_m,
$$
and
$$
\displaystyle\frac{dx_s}{dt}=\sigma(y_s-x_s)+u(x_s-x_m),~~\frac{dy_s}{dt}=rx_s-y_s-x_sz_m+u(y_s-y_m).
$$
Numerical simulations are shown in Fig. 2.

Lastly, we study the R\"{o}ssler system [11],
$$
\displaystyle\frac{dx}{dt}=-y-z,~~ \frac{dy}{dt}=x+ay,~~
\frac{dz}{dt}=b+z(x-c).
$$
This system is dissipative and has a chaotic attractor when
$a=b=0.2$, $c=5.7$. R\"{o}ssler attractor is somewhat difficult to
synchronize by usual methods. Using the present method, this
chaotic attractor can be easily synchronized. Similar to the above
discussion, the drive-response configuration is built as follow,
$$
\displaystyle\frac{dx_m}{dt}=-y_m-z_m,~~\frac{dy_m}{dt}=x_m+ay_m,~~\frac{dz_m}{dt}=b+z_m(x_m-c),
$$
and
$$
\displaystyle\frac{dx_s}{dt}=-y_s-z_s+u(x_s-x_m),~~\frac{dy_s}{dt}=x_s+ay_s+u(y_s-y_m),
~~\frac{dz_s}{dt}=b+z_s(x_s-c)+u(x_s-z_m).
$$
During the process of numerical calculations, we find that chaos
synchronization is reached if we simply choose $u=-6$. These
simulations are presented in Fig. 3.

\newpage
\begin{figure}[!htbp]
\begin{center}
(a)\includegraphics[height=9cm,width=9cm]{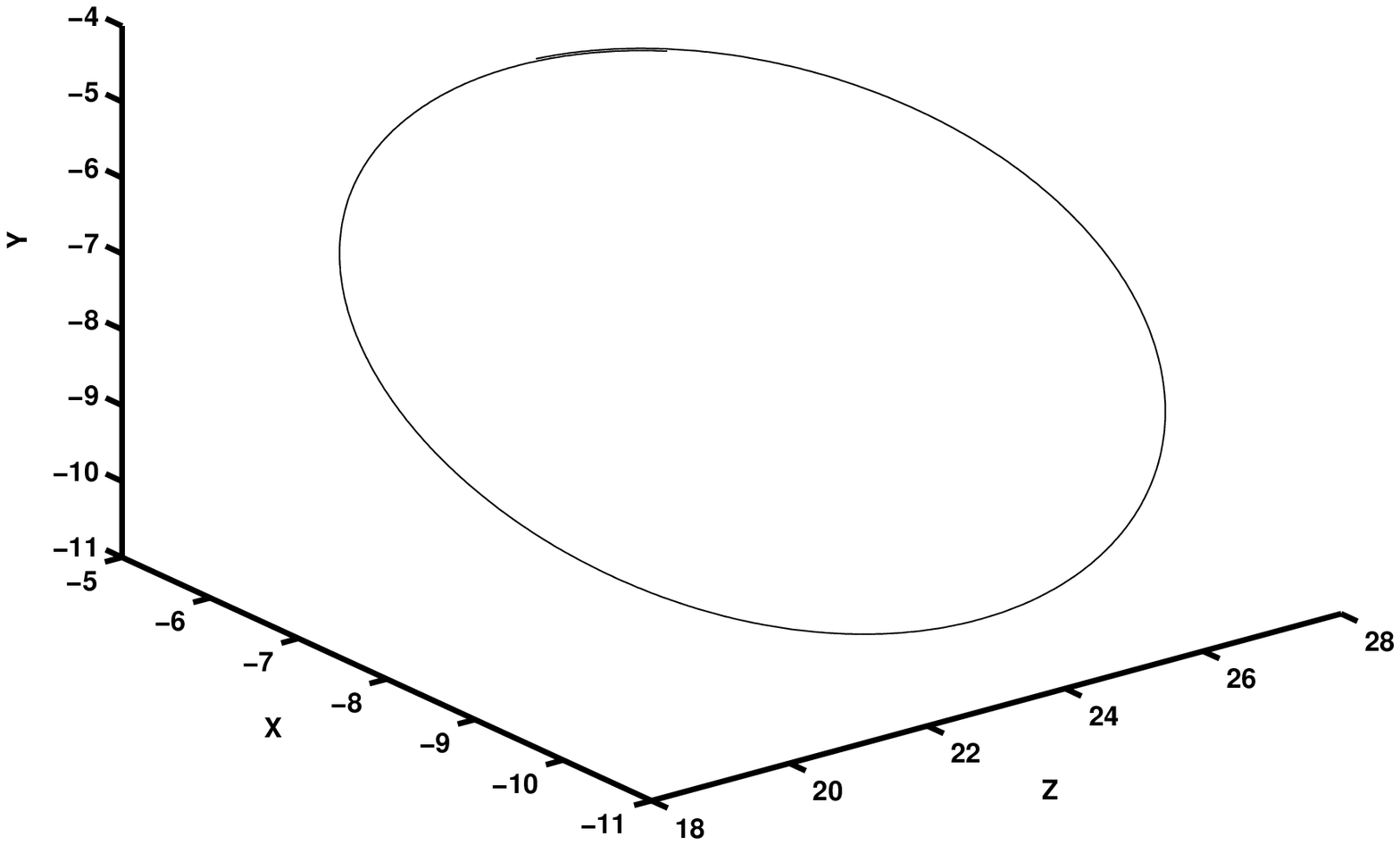}
\vskip 1cm
(b)\includegraphics[height=9cm,width=9cm]{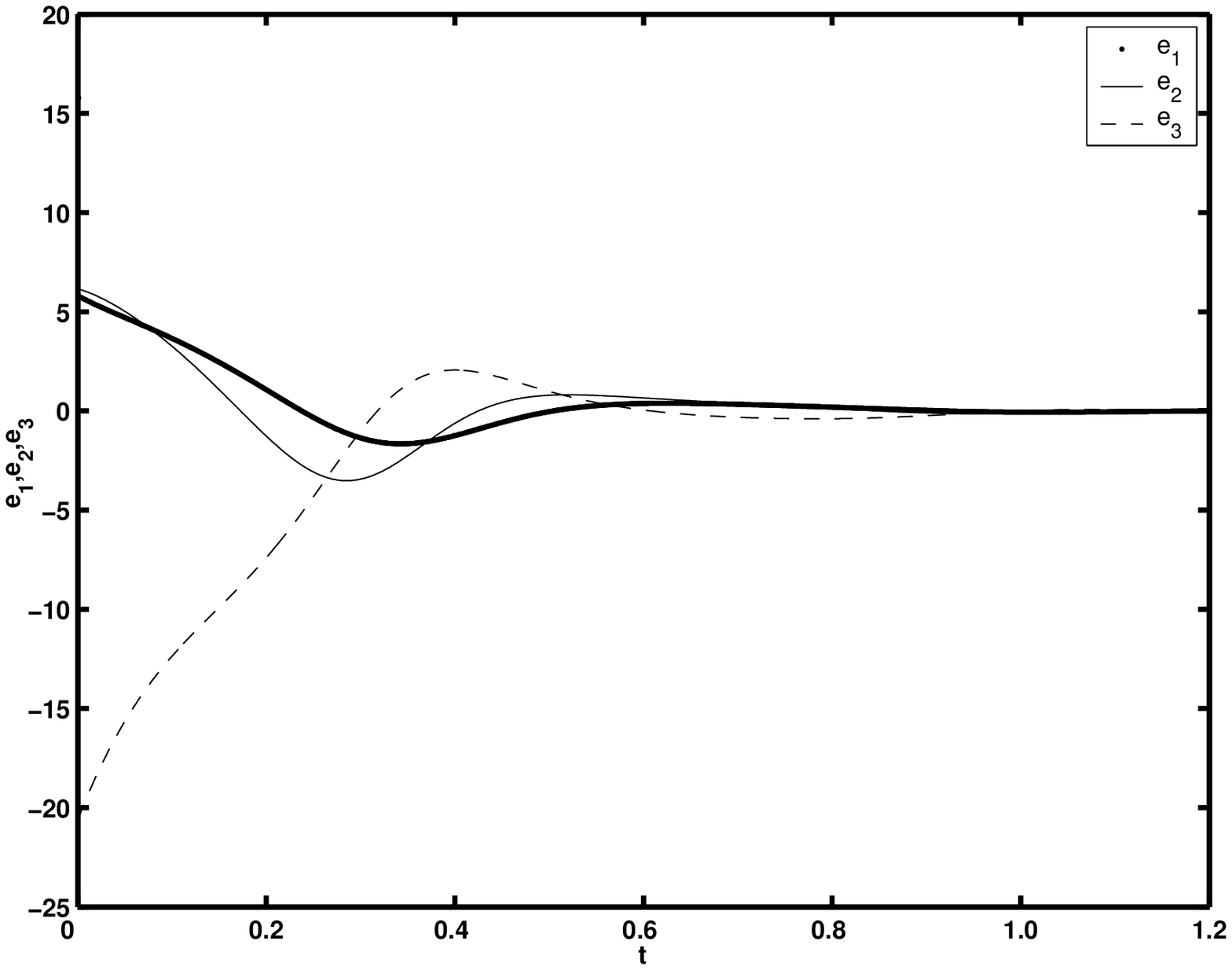}
\caption{Synchronization of unstable limit cycle of Lorenz system.
Here $(\sigma,\,b,\,r)=(10,\,8/3,\,24.5)$, $u=-5$,
 the time step-length is 0.001. (a) The unstable Limit cycle $\mathcal{C}$.
 (b) The evolution of synchronization
error. Here $e_1=x_s-x_m,\, e_2=y_s-y_m,\,e_3=z_s-z_m.$}
\end{center}
\end{figure}

\newpage
\begin{figure}[!htbp]
\begin{center}
(a)\includegraphics[height=9cm,width=9cm]{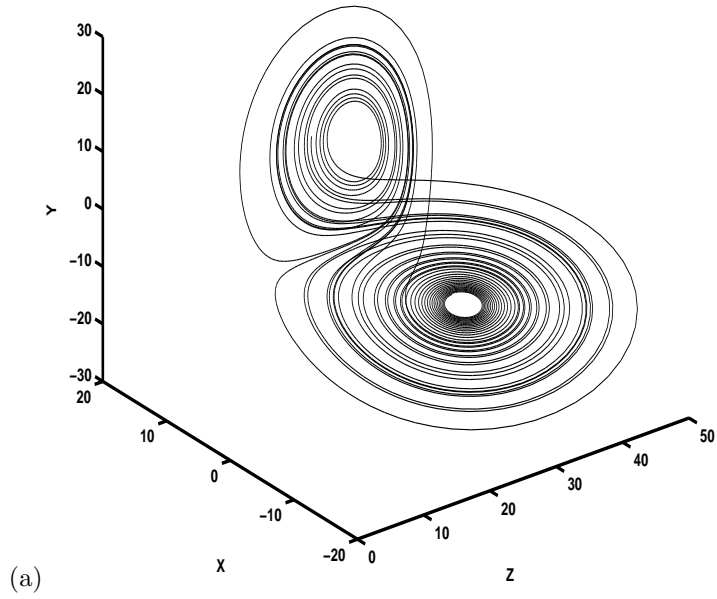}
\vskip 1cm
(b)\includegraphics[height=9cm,width=9cm]{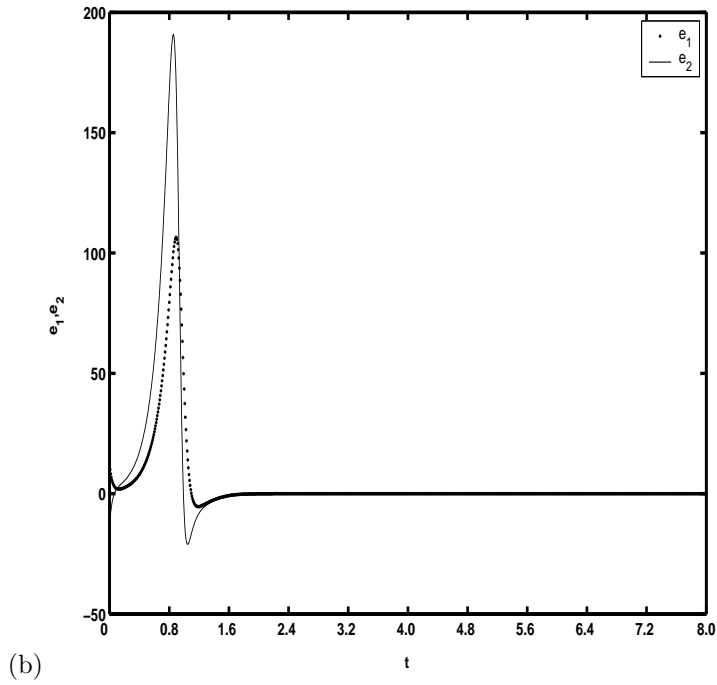}
\caption{Chaos synchronization of Lorenz System. Here
$(\sigma,\,b,\,r)=(10,\,8/3,\,28)$, $u=-6$,
 the time step-length is 0.008. (a) The chaotic attractor.
(b) The evolution of synchronization error. Here,
$e_1=x_s-x_m,\,e_2=y_s-y_m.$}
\end{center}
\end{figure}

\newpage
\begin{figure}[!htbp]
\begin{center}
(a)\includegraphics[height=9cm,width=9cm]{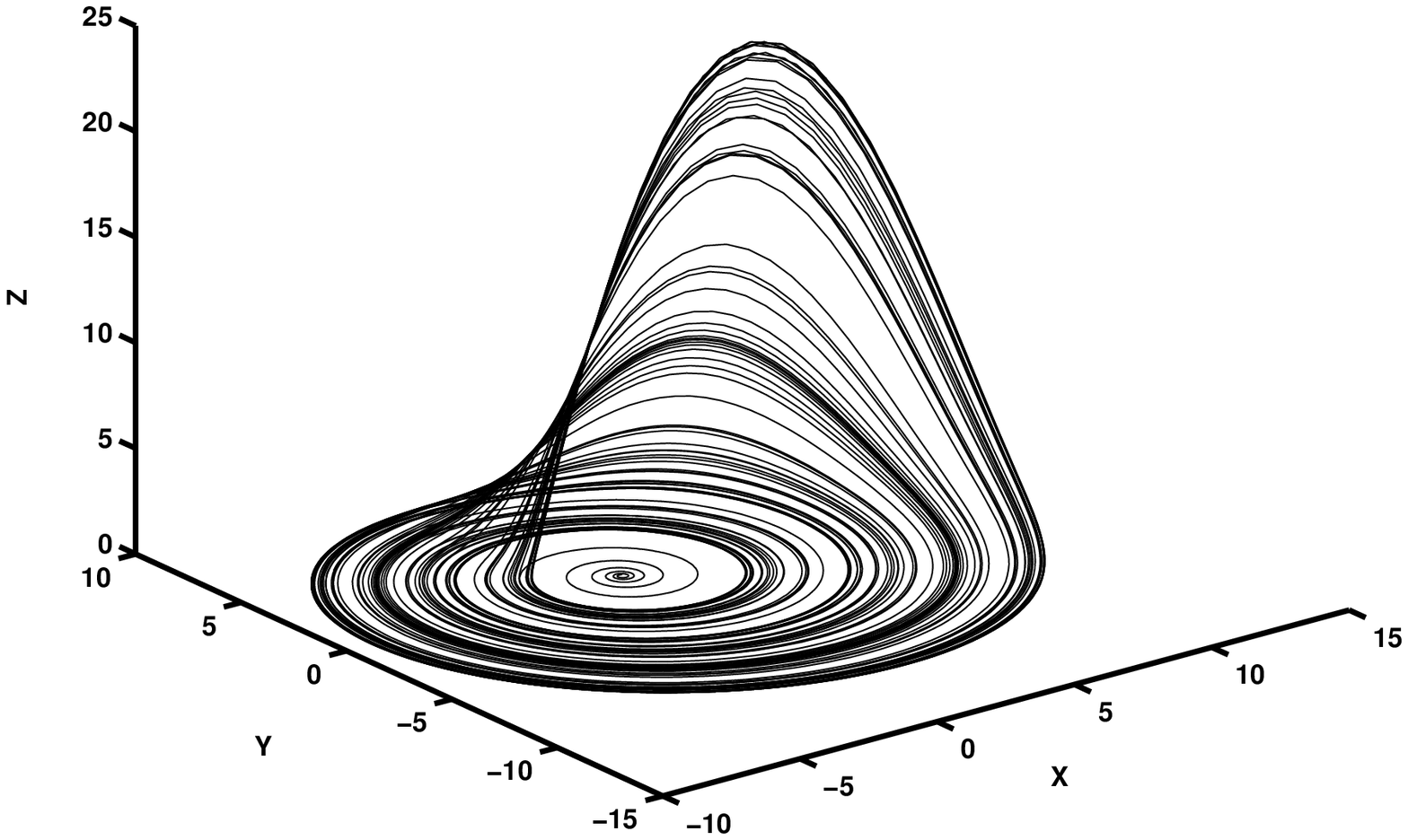}
\vskip 1cm
(b)\includegraphics[height=9cm,width=9cm]{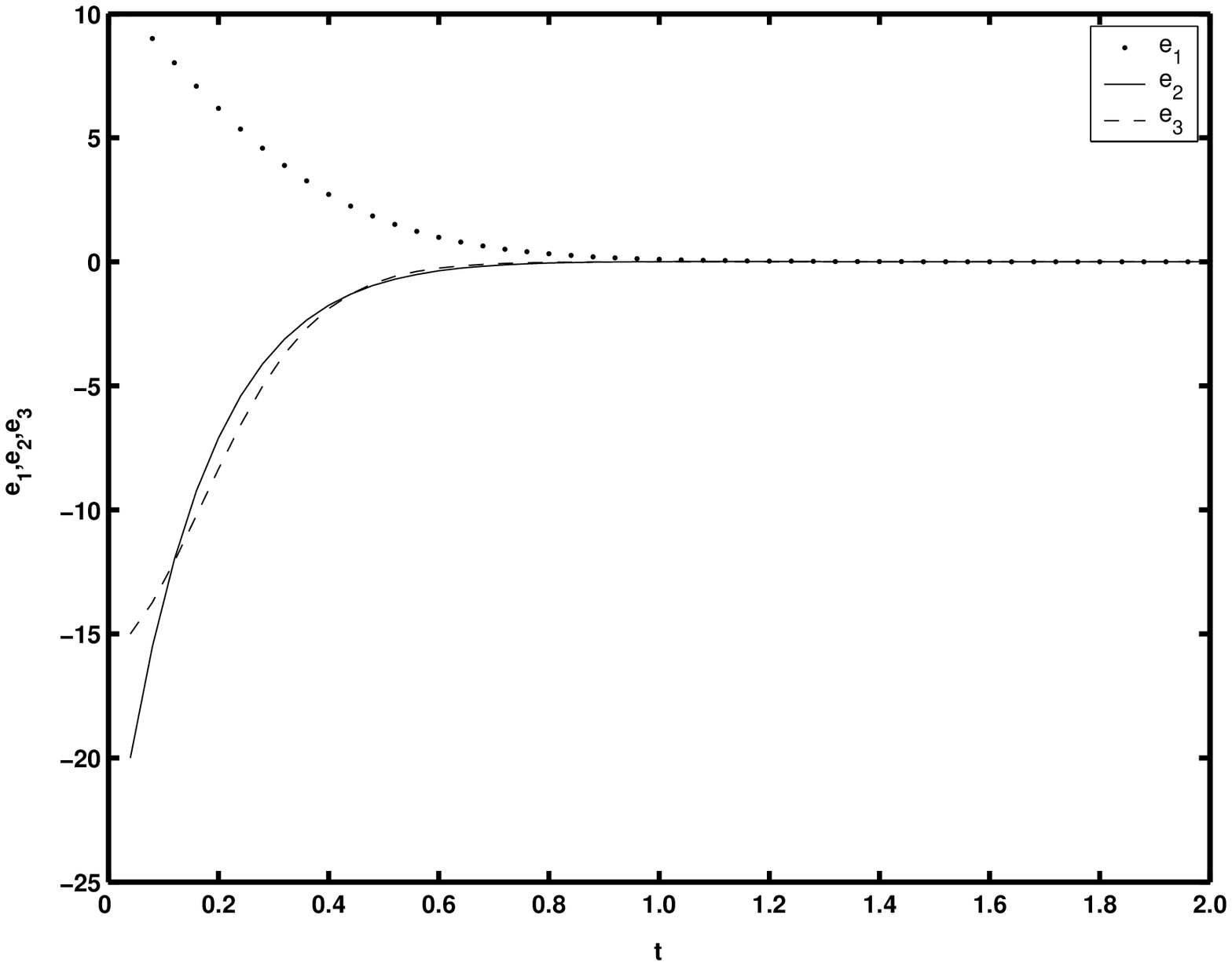}
\caption{Chaos synchronization of R\"{o}ssler System.  Here
$a=b=0.2$, $c=5.7$, $u=-6$, the time step-length is 0.04. (a) The
chaotic attractor.
(b) The evolution of synchronization error. Here,
$e_1=x_s-x_m,\,e_2=y_s-y_m,\,e_3=z_s-z_m.$}
\end{center}
\end{figure}

\bigbreak

\def\item{\par\hang\textindent}
\small
\medskip

\item{[1]} I.I. Blekman, Synchronization in science and
technology, ASME Press, New York, 1988.

\item{[2]} L. M. Pecora, T. L. Carroll, 
Phys. Rev. Lett. {\bf 64}, 821 (1990).

\item{[3]} A. Maybhate, and R.E. Amritkar, Phys. Rev. E {\bf 59},
284 (1999).

\item{[4]} J.W. Shuai, K.W. Wong and L.M. Cheng, Phys. Rev. E {\bf
56}, 2272 (1997).

\item{[5]} C. Zhou and C.H. Lai, Phys. D {\bf 135}, 1 (2000).

\item{[6]} S. Boccaletti, J. Kurths, G. Osipov, D. L.
Valladares, C. S. Zhou, 
Phys. Rep. {\bf 366}, 1 (2002). 

\item{[7]} J. P. Yan and C. P. Li, Chaos, Solitons and Fractals
{\bf 23}, 1683 (2005).

\item{[8]} A. Ambrosetti and G. Prodi, A prime of nonlinear
analysis, Cambridge University Press, Cambridge, 1993.

\item{[9]} C. P. Li and G. Chen, Chaos, Solitons and Fractals {\bf
18}, 69 (2003).

\item{[10]} C. Sparrow, The Lorenz equations, bifurcation, chaos,
and strange attractors, Springer-Verlag, New York, 1982.

\item{[11]} O. E. R\"{o}ssler, Ann. N.Y. Acad. Sci. {\bf 316}, 376
(1978).

\end{document}